\documentclass[11pt,a4paper,twoside,groupcitations]{article}
\usepackage[T1]{fontenc}
\usepackage[ansinew]{inputenc}
\usepackage[english]{babel}
\usepackage{amsfonts}
\usepackage{amsmath}
\usepackage{bm}
\usepackage{array}
\usepackage{amsthm}
\usepackage{amssymb}
\usepackage{graphicx}
\usepackage{subfigure}
\usepackage{braket}
\usepackage{eucal}
\usepackage{verbatim}
\usepackage[table]{xcolor}
\usepackage{caption}
\usepackage{cite}
\usepackage{textcomp}
\usepackage{multicol}
\usepackage{tikz}
\usetikzlibrary{positioning,arrows}
\usetikzlibrary{decorations.pathmorphing}
\usetikzlibrary{decorations.markings}
\usetikzlibrary{calc,decorations.markings}
\usetikzlibrary{arrows,shapes}
\usetikzlibrary{matrix,arrows}
\usepackage{pgfplots}
\usepackage{xparse}
\definecolor{jade}{HTML}{00A86B}
\newcommand{\be}{\begin{eqnarray}}
\newcommand{\ee}{\end{eqnarray}}

\newcommand{\expec}[1]{\mbox{$\langle\, #1\,\rangle$}}

\renewcommand{\a}{\hat a}
\newcommand{\ac}{\hat a^{\dagger}}

\renewcommand{\d}{\mbox{${\rm d}$}} 
\newcommand{\lp}{\ell_{\rm p}}
\newcommand{\mpl}{m_{\rm p}}
\newcommand{\gn}{G_{\rm N}}

\newcommand{\Rh}{R_{\rm H}}

\newcommand{\jb}{J_{\rm B}}

\newcommand{\Ng}{{N_{\rm G}}}
\newcommand{\Nge}{{N^{\rm eff}_{\rm G}}}
\newcommand{\Nb}{{N_{\rm B}}}
\newcommand{\mub}{{\mu_{\rm B}}}
\newcommand{\Rinf}{{R_{\infty}}}
\title{\bf Quantum black holes in bootstrapped Newtonian gravity}
\author{Roberto~Casadio$^{ab}$\thanks{E-mail: casadio@bo.infn.it},
$\ $
Michele~Lenzi$^{ab}$\thanks{E-mail: michele.lenzi2@unibo.it}
$\ $
and 
Alessandro~Ciarfella$^{a}$\thanks{E-mail: alessandro.ciarfella@studio.unibo.it}
\\
\\
$^a${\em Dipartimento di Fisica e Astronomia, Universit\`a di Bologna}
\\
{\em via Irnerio~46, 40126 Bologna, Italy}
\\
\\
$^b${\em I.N.F.N., Sezione di Bologna, I.S.~FLAG}
\\
{\em viale B.~Pichat~6/2, 40127 Bologna, Italy}
}
\begin{document}
\maketitle
\begin{abstract}
We analyse the classical configurations of a bootstrapped Newtonian potential generated by
homogeneous spherically symmetric sources in terms of a quantum coherent state.
We first compute how the mass and mean wavelength of these solutions scale in terms
of the number of quanta in the coherent state.
We then note that the classical relation between the ADM mass and the proper mass of the source
naturally gives rise to a Generalised Uncertainty Principle for the size of the gravitational radius
in the quantum theory.
Consistency of the mass and wavelength scalings with this GUP requires the compactness remains
at most of order one even for black holes, and the corpuscular predictions are thus recovered,
with the quantised horizon area expressed in terms of the number of quanta in the
coherent state.
Our findings could be useful for analysing the classicalization of gravity in the presence of matter
and the avoidance of singularities in the gravitational collapse of compact sources.
\par
\null
\par
\noindent
\textit{PACS - 04.70.Dy, 04.70.-s, 04.60.-m}
\end{abstract}
\section{Introduction and motivation}
\setcounter{equation}{0}
\label{Sintro}
Black holes represent a benchmark for any attempt at quantising gravity.
According to general relativity, the gravitational collapse of any compact source will generate geodetically
incomplete space-times if a trapping surface appears~\cite{HE},
whereas an eternal point-like source is mathematically incompatible with the Einstein field
equations~\cite{geroch}.
We expect a quantum theory of the gravitational interaction should fix this inconsistent classical picture, 
like quantum mechanics explains the stability of the hydrogen atom.
Whether this can be achieved by modifications of the gravitational dynamics solely at the Planck scale or 
with sizeable implications for macroscopic phenomenology remains open to debate.
\par
The recently proposed corpuscular model of black holes~\cite{DvaliGomez} abandons the geometric interpretation 
of gravity at the root of general relativity and belongs to the class of approaches for which geometry
should only emerge at suitable (macroscopic) scales from the underlying (microscopic) quantum field theory of
gravitons.
It is in particular based on the idea that the constituents of black holes are soft gravitons (marginally) bound in their
own potential and forming a condensate~\cite{DvaliGomez,germani}.
The characteristic Compton-de~Broglie wavelength of these gravitons should be 
\be
\lambda_{\rm G}\sim \Rh
\ ,
\label{lambdaRh}
\ee
where the (gravitational) Schwarzschild radius of the black hole of Arnowitt-Deser-Misner (ADM) mass~\cite{ADM}
$M$ is given by~\footnote{We shall use units with $c=1$ and the Newton constant $\gn=\lp/\mpl$, where $\lp$
is the Planck length and $\mpl$ the Planck mass (so that $\hbar=\lp\,\mpl$).}
\be
\Rh=2\,\gn\,M
\ ,
\label{Rh}
\ee
and the energy of the gravitons is correspondingly given by $\epsilon_{\rm G}\sim \hbar/\lambda_{\rm G}$.
If one assumes that the total mass of the black hole $M\simeq N_{\rm G}\,\epsilon_{\rm G}$, there immediately
follows the scaling relation 
\be
N_{\rm G}
\sim 
\frac{M^2}{\mpl^2}
\sim
\frac{\Rh^2}{\lp^2}
\ ,
\label{Ng}
\ee
a result which reproduces Bekenstein's conjecture for the quantisation of the horizon area~\cite{bekenstein},
and indeed holds for any compact sources, as we shall review in the following.
\par
Black hole formation by gravitational collapse requires the presence of matter 
in any astrophysically realistic situations,~\footnote{Of course, one could also envisage the creation
of black holes by focusing gravitational waves, but highly energetic processes involving matter would presumably
be needed in order to produce those waves in the first place.}
whose inclusion then
allows for looking at a connection with the post-Newtonian approximation~\cite{Casadio:2016zpl}.
This can be seen by considering that the (negative) gravitational energy of a source of mass $M$
localised inside a sphere of radius $R$ is given by
\be
U_{\rm N}
\sim
M\, V_{\rm N}(R)
\sim
-\frac{\gn\,M^2}{R}
\ ,
\ee
where $V_{\rm N}\sim -{\gn\,M}/{r}$ is the (negative) Newtonian potential.
This classical potential can be reproduced by the expectation value of a scalar field on a coherent state $\ket{g}$,
whose normalisation then yields the graviton
number~\eqref{Ng}~\cite{Casadio:2016zpl,Casadio:2017cdv,Mueck:2013mha}.
In addition to that, assuming most gravitons have the same wave-length $\lambda_{\rm G}$,
the (negative) energy of each single graviton is correspondingly given by
\be
\epsilon_{\rm G}
\sim
\frac{U_{\rm N}}{N_{\rm G}}
\sim
-\frac{\lp\,\mpl}{R} 
\ ,
\ee
which yields the typical Compton-de~Broglie length $\lambda_{\rm G}\sim R$.
The graviton self-interaction energy hence reproduces the (positive) post-Newtonian energy,
\be
U_{\rm GG}(R)
\sim
N_{\rm G} \, \epsilon_{\rm G} \, V_{\rm N}(R)
\sim
\frac{\gn^2\,M^3}{R^2}
\ ,
\ee
and the fact that gravitons in a black hole are marginally bound is reflected by the {\em maximal
packing} condition~\cite{DvaliGomez}, which roughly reads $U_{\rm N}+U_{\rm GG}\simeq 0$
for $R\simeq \Rh$~\cite{Casadio:2016zpl,Casadio:2017cdv}.
\par
Small (post-Newtonian) perturbations around the Newtonian potential were analysed in more details
in Ref.~\cite{Casadio:2017cdv}.
However, since the post-Newtonian correction $V_{\rm PN}\sim 1/r^2$ is positive and grows faster than
the Newtonian potential closer to the surface of the source, one cannot consider matter sources with
radius $R\lesssim \Rh$ in this approximation.
For that purpose, a bootstrapped Newtonian potential $V$ satisfying a nonlinear equation for a spherically
symmetric and static source was derived in Ref.~\cite{Casadio:2017cdv} and
subsequently studied~\cite{BootN} and improved~\cite{Casadio:2019cux}. 
The final form of the governing equation contains, besides the usual Newtonian coupling with the matter density,
a coupling with the internal pressure and a gravitational self-interaction term, all of which are treated
non-perturbatively on the same footing in order to explore the effects of nonlinearities in the strong field
regime~\cite{Casadio:2019cux,Casadio:2019pli}.
Solutions were found for uniform sources of proper mass $M_0$ with generic compactness $\gn\,M/R\sim \Rh/R$,
from the weak field regime $R\gg\Rh$, in which we recover the standard post-Newtonian picture with $M\simeq M_0$,
to the large compactness case $R\lesssim\Rh$ where we find the proper mass $M_0$ significantly differs from
the ADM mass $M$ and the source is enclosed within a (Newtonian) horizon~\cite{Casadio:2019cux}.
It is the latter case which we can naively view as describing black holes in bootstrapped Newtonian gravity,
and it is natural to ask if quantum effects could imply a constraint on the maximum compactness of the source
in order to recover the maximal packing mentioned above.
\par
Like the Newtonian analogue, the bootstrapped potential determines the gravitational pull acting on
test particles at rest.~\footnote{In a quantum field theory description, this dynamics would be obtained
from transition amplitudes yielding the propagator of the test particle.
We here assume that all the required approximations leading to the effective appearance of a potential
hold.}
It can therefore be used in order to describe the mean field force acting on the constituents of the
system, namely the baryons in the static matter source as well as the gravitons in the potential itself.
In order to gain some insight into the quantum structure of such self-gravitating systems, the solutions for the
bootstrapped potential will be here described in terms of the quantum coherent state of a free massless scalar field,
analogously to what was done for the Newtonian potential in Ref.~\cite{Casadio:2017cdv}
(see also Ref.~\cite{Mueck:2013mha} for a model of black holes, Ref.~\cite{DvaliSoliton}
for general solitons and Refs.~\cite{Mueck:2013wba} for photons in a static electric or magnetic field).
This analysis will be carried out in details both in the Newtonian approximation, which corresponds to sources of
small compactness, and for the large compactness case.
The analysis of the coherent state will allow us to recover the scaling~\eqref{Ng} for the ADM mass $M$
in terms of the number of gravitons $N_{\rm G}$ in all cases, whereas the scaling~\eqref{lambdaRh} for the
mean wavelength will appear to require the fine-tuned maximal packing $R\sim\Rh$.
However, by considering the quantum nature of the source in rather general terms, we will also find that the
classical bootstrapped relation between the black hole mass $M$ and the proper mass $M_0$ of the source implies  
a Generalised Uncertainty Principle (GUP)~\cite{pGUP} for the horizon size.
Moreover, consistency of this GUP with the properties of the coherent state indeed suggests that
the compactness of the source should be at most of order one and the scaling relation~\eqref{lambdaRh}
can therefore be recovered in a fully quantum description of black holes.
Such a bound on the maximum compactness of self-gravitating objects is at the heart of the so
called {\em classicalization\/} of gravity~\cite{classicalization}, according to which quantum fluctuations
involved in processes above the Planck scale should be suppressed precisely by the
formation of black holes viewed as quasi-classical configurations.
\par
The paper is organised as follows:
in the next Section, we review the coherent state description for a static potential and apply it to 
the Newtonian potential generated by a uniform source;
in Section~\ref{boot}, we recall the fundamentals of  the bootstrapped Newtonian picture,
for which we then repeat the analysis in terms of a coherent state in Section~\ref{scaling}
(with more technical details given in Appendix~\ref{calculations}).
In that Section, we will derive the main results mentioned above, with final comments
and outlook in Section~\ref{s:conc}.
\section{Quantum coherent state}
\label{coherent}
\setcounter{equation}{0}
We will first review how to describe a generic static potential $V$ by means of the coherent state of
a free massless scalar field.
This will allow us to introduce a formal way of counting the number of quanta $N_{\rm G}$ for any such potential.
We remark that a clear understanding of the physical meaning of the number of quanta so defined, in a field configuration
that is not  in general perturbatively related with the vacuum, could possibly be obtained only by
studying the dynamical process leading to the formation of such a configuration.
Of course, there is little hope of solving this problem analytically in a non-linear theory.
Like in Refs.~\cite{Casadio:2016zpl,Casadio:2017cdv}, we shall instead take a similar approach to that for general solitons
in quantum field theory found in Ref.~\cite{DvaliSoliton}
(see also Ref.~\cite{Mueck:2013mha} for a model of black holes and Refs.~\cite{Mueck:2013wba} for photons in QED).
We remark, in fact, that for our purposes, the number $N_{\rm G}$ is mostly an auxiliary quantity which allows us to tackle the
issue of classicalization by means of the corresponding scaling relations~\eqref{lambdaRh} and~\eqref{Ng}
for black holes, as discussed in Section~\ref{Sintro}.
\par
We start by setting the stage for the quantum interpretation of the dimensionless $V=V(\bm{x})$
based on simple Fourier transforms.
In order to fix the notation, we write normalised plane waves in the three-dimensional space
$\mathbb{R}^3=\{{\bm{x}}=(x^1,x^2,x^3):\, x^i\in\mathbb{R}\}$ as
\be
\label{vk}
v_{\bm{k}}(\bm{x}) 
\equiv
\frac{e^{i\,\bm{k}\cdot\bm{x}}}{(2\,\pi)^{3/2}}
\ ,
\ee
so that they satisfy the orthogonality relation
\be \label{ortovk}
\int_{\mathbb{R}^3}
\d\bm{x}\,
v^*_{\bm{k}}(\bm{x})\, v_{\bm{h}}(\bm{x})
=
\delta(\bm{k}-\bm{h})
\ .
\ee
We can then expand the real potential as
\be
\label{antifourierV}
V(\bm{x}) 
=
\int_{\mathbb{R}^3} 
\frac{\d\bm{k}}{(2\,\pi)^{3}} 
\,\tilde{V}(\bm{k})\,
v_{\bm{k}}(\bm{x})
\ ,
\ee
where, in turn, one has 
\be
\label{fourierV}
\tilde{V}(\bm{k})
=
\int_{\mathbb{R}^3}
\d\bm{x} 
\,V(\bm{x})\,v^*_{\bm{k}}(\bm{x})
\ ,
\ee
with $\tilde V(\bm{k})=\tilde V^*(-\bm{k})$.
\par
Next, we will specialise to spherically symmetric cases and apply the construction to the Newtonian
potential generated by a uniform ball of matter, for which the Fourier transform can be computed
explicitly.~\footnote{The even simpler cases of the Newtonian potential for a point-like source and
for a Gaussian source can be found, {\em e.g.}~in Ref.~\cite{Casadio:2017cdv}.}
This exercise will allow us to introduce in the next Section a different way of analysing cases,
like the bootstrapped Newtonian potential, for which this cannot be done 
analytically. 
\subsection{Static scalar potential}
As it was done in Ref.~\cite{Casadio:2017cdv}, the first step consists in rescaling the potential $V$ so as
to obtain a canonically normalised real scalar field~\footnote{We recall that a canonically normalised scalar
field has dimensions of $\sqrt{\rm{mass/length}}$.}
\be
\label{phiV}
\Phi 
=
\sqrt{\frac{\mpl}{\lp}}\, V
\ .
\ee
We will then quantise $\Phi$ as a free massless field satisfying the wave equation
\be
\left(-\partial_t^2+\partial_{x^1}^2+\partial_{x^2}^2+\partial_{x^3}^2\right)
\Phi(t,\bm x)
\equiv
\left(-\partial_t^2+\triangle\right)
\Phi
=
0
\ ,
\ee
whose solutions are given by
\be
\label{uk}
u_{\bm{k}}(t,\bm{x}) 
= 
v_{\bm{k}}(\bm{x})\, 
e^{-i\,k\,t}
\ ,
\ee
with $k=\sqrt{\bm{k}\cdot\bm{k}}$, and satisfy the orthogonality relation in the Klein-Gordon
scalar product~\footnote{We will usually omit the domain of integration when it is given by all
of ${\mathbb{R}^3}$.}
\be
i
\int
\d\bm{x}
\left[ 
u^*_{\bm{k}}(t,\bm{x})\partial_t 
u_{\bm{h}}(t,\bm{x})
-
\partial_t
u^*_{\bm{k}}(t,\bm{x})\,
u_{\bm{h}}(t,\bm{x})
\right]
= 
\delta(\bm{k}-\bm{h})
\ .
\ee
The quantum field operator and its conjugate momentum then read
\be
\label{Phi}
\hat{\Phi}(t,\bm{x})
\!\!&=&\!\! 
\int
\frac{\d\bm{k}}{(2\,\pi)^{3}} 
\sqrt{\frac{\lp\,\mpl}{2\,k}} 
\left(
\a_{\bm{k}}\,
e^{-i\,k\,t + i\,\bm{k}\cdot\bm{x}}
+
\ac_{\bm{k}}\, e^{i\,k\,t - i\,\bm{k}\cdot\bm{x}}\right) 
\\
\hat{\Pi}(t,\bm{x}) 
\!\!&=&\!\!
i\int \frac{\d\bm{k}}{(2\,\pi)^{3}}  
\sqrt{\frac{\lp\,\mpl\,k}{2}} 
\left( - \a_{\bm{k}}\,e^{-i\,k\,t
+
i\,\bm{k}\cdot\bm{x}} 
+
\ac_{\bm{k}}\,
e^{i\,k\,t - i\,\bm{k}\cdot\bm{x}}
\right)
\ ,
\ee
and must satisfy the equal time commutation relations
\be
\left[\hat{\Phi}(t,\bm{x})
,\hat{\Pi}(t,\bm{y})\right] 
=
i\,\hbar\,\delta(\bm{x}-\bm{y})
\ .
\ee
The creation and annihilation operators therefore obey the standard commutation rules
\be
\left[\a_{\bm{k}},\ac_{\bm{p}}\right]
=
\delta({\bm{k}-\bm{p}})
\ ,
\ee
and the Fock space of quantum states is built from the vacuum $\a_{\bm{k}}\ket{0}=0$.
\par
Classical configurations of the scalar field must be given by suitable states in the Fock space,
and we note that a natural choice for $V=V(\bm{x})$ is given by a coherent state,
\be
\label{akg}
\a_{\bm{k}}
\ket{g} 
=
g_{\bm{k}}\,e^{i\,\gamma_{\bm{k}}(t)}
\ket{g}
\ ,
\ee
such that the expectation value of the quantum field $\hat{\Phi}$ reproduces the classical
potential, namely
\be
\label{expecphi}
\sqrt{\frac{\lp}{\mpl}}
\bra{g}\hat{\Phi}(t,\bm{x})\ket{g}
=
V(\bm{x})
\ .
\ee
From the expansion~\eqref{Phi}, one can easily compute the left hand side of Eq.~\eqref{expecphi} by making use
of Eq.~\eqref{akg}.
Comparing with Eq.~\eqref{antifourierV} then yields
\be
\label{gkgen}
g_{\bm{k}}
=
\frac{1}{\lp}\,\sqrt{\frac{k}{2}}\,\tilde{V}(\bm{k})
\ee
and $\gamma_{\bm{k}}(t) = k\,t$, with the latter condition turning (propagating) plane waves into standing waves.
\par 
We are particularly interested in the total number of quanta in this coherent state,
whose general expression is given by
\be 
N
\!\!&=&\!\!
\int 
\frac{\d\bm{k}}{\left(2\,\pi \right)^3} 
\bra{g}
\ac_{\bm{k}}\, \a_{\bm{k}}
\ket{g}
\nonumber
\\
\!\!&=&\!\!
\int 
\frac{\d\bm{k}}{\left(2\,\pi \right)^3} 
\,g_{\bm{k}}^2
\nonumber
\\
\!\!&=&\!\!
\frac{1}{2\,\lp^2}
\int 
\frac{\d\bm{k}}{\left(2\,\pi \right)^3} 
\,k\,\tilde{V}^2(\bm{k})
\ ,
\label{NG0}
\ee
and in their mean wavelength $\lambda\simeq 1/\bar{k}\equiv N/\expec{k}$, where the mean
wavenumber is given by
\be
\expec{k}
\!\!&=&\!\! 
\int 
\frac{\d\bm{k}}{\left(2\,\pi \right)^3} 
\bra{g}\,k\,\ac_{\bm{k}}\,\a_{\bm{k}}\,\ket{g}
\nonumber
\\
\!\!&=&\!\!
\int 
\frac{\d\bm{k}}{\left(2\,\pi \right)^3} \,
k\,g_{\bm{k}}^2
\nonumber
\\
\!\!&=&\!\! 
\frac{1}{2\,\lp^2}
\int 
\frac{\d\bm{k}}{\left(2\,\pi \right)^3} 
\,k^2\,\tilde{V} ^2(\bm{k})
\ .
\label{kmed0}
\ee
The above general expressions will next be specified for the Newtonian potential generated 
by spherically symmetric sources.
\subsection{Newtonian potential for spherical sources}
The Newtonian potential $V(\bm{x}) = V_{\rm N}(r)$ for a spherically symmetric source of static energy density
$\rho=\rho(r)$, can be described by means of the Lagrangian
\be
L_{\rm N}[V_{\rm N}]
=
-4\,\pi
\int_0^\infty
r^2 \,\d r
\left[
\frac{\left(V_{\rm N}'\right)^2}{8\,\pi\,\gn}
+\rho\,V_{\rm N}
\right]
\ ,
\label{LagrNewt}
\ee
where $f'\equiv\d f/\d r$.
The corresponding Euler-Lagrange equation of motion is the Poisson equation in spherical coordinates,
\be
r^{-2}\left(r^2\,V_{\rm N}'\right)'
\equiv
\triangle V_{\rm N}
=
4\,\pi\,\gn\,\rho
\label{EOMn}
\ .
\ee
Since the system is static, the (on-shell) Hamiltonian is simply given by $H_{\rm N}[V_{\rm N}]=-L_{\rm N}[V_{\rm N}]$.
After introducing the rescaled field $\Phi$ of Eq.~\eqref{phiV}, we also need to rescale the Hamiltonian by a factor
of $4\,\pi$ in order to canonically normalise the kinetic term~\footnote{See Ref.~\cite{Casadio:2017cdv} for more details.}, 
to wit
\be
\label{HN}
H_{\rm N}[\Phi]
=
4\,\pi\,
H_{\rm N}[V_{\rm N}]
\ .
\ee 
The previous general analysis for the coherent state can now be adapted to the spherically symmetric case by just
replacing the plane waves~\eqref{vk} withe spherical Bessel functions~\cite{Casadio:2017cdv}, 
\be
v_{\bm{k}}(\bm{x})
\to
j_0(k\,R)
\equiv
\frac{\sin(k\,R)}{k\,R}
\ .
\ee 
By substituting Eq.~\eqref{antifourierV} into Eq.~\eqref{EOMn}, we obtain the general result
\be
\tilde{V}_{\rm N}(k)
=
-
\frac{4\,\pi\,\lp\,\tilde{\rho}(k)}{\mpl\,k^2}
\ ,
\ee
which, together with Eq.~\eqref{gkgen}, leads to
\be
g_k
=
-
\frac{4\,\pi\,\tilde{\rho}(k)}
{\mpl\sqrt{2\,k^3}}
\ .
\ee
The spherically symmetric versions of Eqs.~\eqref{NG0} and~\eqref{kmed0} then read
\be
\label{NGN}
\Ng
=
\int_0^{\infty} 
\frac{\d k}{2\,\pi^2} 
\,k^2\,g_k ^2
\ ,
\ee
and 
\be
\label{EN}
\expec{k}
=
\int_0^{\infty} 
\frac{\d k}{2\,\pi ^2} 
\,k^3\,g_k ^2
\ ,
\ee
where the suffix $\rm G$ emphasises that the quantity is evaluated 
in the coherent state representing the gravitational potential.
\subsection{Newtonian potential of a uniform ball}
\label{Newt}
Note that all expressions above can be explicitly computed if we know the coefficients $g_k$.
As a workable example, we will consider a homogeneous source of radius $R$, whose density is given by
\be
\rho
=
\rho_0
\equiv
\frac{3\, M_0}{4\,\pi\, R^3}\, 
\Theta(R-r)
\ ,
\label{HomDens}
\ee
where 
\be
M_0
=
4\,\pi
\int_0^R
\d r\,r^2\,\rho(r)
\simeq
\Nb\,\mub
\ee
is the total rest mass of the homogeneous configuration of $\Nb$ baryonic constituents with proper mass $\mub$.
The solution to Eq.~\eqref{EOMn} must satisfy the regularity condition in the origin
\be
\label{bound0}
V_{\rm in}'(0)
=
0 
\ ,
\ee
where $V_{\rm in}=V_{\rm N}(r<R)$,
and it must also be smooth across the surface $r=R$, 
\be
\left\{
\begin{array}{l}
V_{\rm in}(R)
=
V_{\rm out}(R)
\equiv
V_R
\\
\\ 
V'_{\rm in}(R)
=
V'_{\rm out}(R)
\equiv 
V'_R 
\ ,
\end{array}
\right.
\label{boundR}
\ee
where $V_{\rm out}=V_{\rm N}(r>R)$.
The complete solution is in fact well-known and reads
\be
\label{VN}
V_{\rm N}
=
\left\{
\begin{array}{lrl}
\strut\displaystyle\frac{\gn\, M}{2\,R^3}\left(r^2 - 3\,R^2\right)
&
{\rm for}
&
0\le  r<R 
\\
\\
-\strut\displaystyle\frac{\gn\,M}{r}
&
{\rm for}
&
r>R
\ ,
\end{array}
\right.
\ee
where $M=M_0$ is the ADM mass equal to the rest mass in this Newtonian case.
\par 
The Fourier transform of the density~\eqref{HomDens} is given by
\be
\tilde{\rho}(k)
=
4\,\pi\int_0^{\infty}
\d r\,r^2\,\rho(r)\,
j_0(k\,r)
=
\frac{3\,M}{k^2\,R^2}
\left[\frac{\sin(k\,R)}{k\,R}
- \cos(k\,R) \right]
\ ,
\ee
and the coherent state eigenvalues then read
\be \label{gkN}
g_k
=
\frac{12\,\pi\,M}{\sqrt{2}\,\mpl\,k^{7/2}\,R^2}
\left[
\cos{(k\,R)}
-
\frac{\sin{(k\,R)}}{k\,R}
\right]
\ .
\ee
The mean wavenumber~\eqref{EN} can be easily evaluated from this expression,
\be
\notag
\expec{k}
&\!\!=\!\!&
\frac{36\,M^2}
{\mpl^2\,R^4}
\int_{0}^{\infty}
\frac{\d k}{k^4}
\left[
\cos{(k\,R)}
-
\frac{\sin{(k\,R)}}{k\,R}
\right]^2
\\ 
\notag
&\!\!=\!\!&
\frac{36\,M^2}
{\mpl^2\,R}
\int_{0}^{\infty}
\frac{\d z}{z^4}
\left[
\cos{z}
-
\frac{\sin{z}}{z}
\right]^2
\\
\label{kN}
&\!\!=\!\!&
\frac{12\,\pi\,M^2}{5\,\mpl^2\,R}
=
-4\,\pi\,\frac{U_{\rm N}}{\lp\,\mpl}
\ ,
\ee
where 
\be
U_{\rm N}
=
-\frac{3\,\gn\,M^2}{5\,R}
\ee 
is precisely the gravitational potential energy of the spherically symmetric homogeneous
source~\eqref{HomDens}, a result consistent with the linearity of the Newtonian interaction~\footnote{We
note that the factor of $4\,\pi$ in the right hand side of Eq.~\eqref{kN} is just a consequence of the canonical
rescaling~\eqref{HN}.}.
\par
While the mean wave number $\expec{k}$ above is finite, the number of gravitons~\eqref{NGN} diverges
in the infrared (IF), {\em i.e.}~$k^2\,g_k^2 \to \infty$ for $k \to 0$.
This is also expected as the potential~\eqref{VN} has infinite spatial support and we could simply introduce a
cut-off $k_0 = 1/\Rinf$ to account for the necessarily finite life-time of a realistic source~\cite{Casadio:2017cdv}.
In this case, 
\be
\Ng
&\!\!=\!\!&
\frac{36\,M^2}
{\mpl^2\,R^4}
\int_{k_0}^\infty
\frac{\d k}{k^5}
\left[
\cos{(k\,R)}
-
\frac{\sin{(k\,R)}}{k\,R}
\right]^2
\notag
\\
&\!\!=\!\!&
\frac{36\,M^2}
{\mpl^2}
\int_{R/\Rinf}^{\infty}
\frac{\d z}{z^5}
\left[
\cos{z}
-
\frac{\sin{z}}{z}
\right]^2
\\
&\!\!\simeq\!\!&
4\,\frac{M^2}{\mpl^2}
\log{\left(\frac{\Rinf}{2\,R}\right)}
\label{NgN}
\ .
\ee
The corpuscular scaling~\eqref{Ng} with the square of the energy $M$ of the system  
already appears at this stage, but we can still understand better the logarithmic divergence for $R_\infty\to \infty$
in order to make full sense of it.
\par
As pointed out in Ref.~\cite{DvaliSoliton}, the fact that the energy (or the mean wavenumber) is finite despite
the diverging number of constituents is a direct consequence of a decreasing energy contribution coming
from gravitons with lower and lower momenta.
We can in fact separate two contributions by introducing a scale $\Lambda$ which splits 
the phase space of gravitons into {\em effective\/} (hard) and IR (soft) modes,
\be
\expec{k}
&\!\!=\!\!&
\int_0^{\Lambda} 
\frac{\d k}{2\,\pi ^2} 
\,k^3\,g_k ^2
+
\int_{\Lambda}^{\infty} 
\frac{\d k}{2\,\pi ^2} 
\,k^3\,g_k ^2
\notag
\\ 
&\!\!\equiv\!\!&
k_{\rm IR} + k_{\rm eff}
\ ,
\label{ksplit}
\ee
where we require $k_{\rm eff}(\Lambda)\gg k_{\rm IR}(\Lambda)$.
Indeed the scale $\Lambda$ remains somewhat arbitrary, since it is just defined by requiring that
$k_{\rm eff}(\Lambda)\simeq \expec{k}$ to a good approximation. 
The accuracy of the approximation is clearly measured by the ratio $k_{\rm IR}/k_{\rm eff}$ which
we plot in Fig.~\ref{ratiok} (see Appendix~\ref{calcfg} for the details).
The interesting fact it that we can identify a threshold value $\Lambda_R\simeq 1/R$
which only depends on the size $R$ of the source and not on $M$.
Values of $\Lambda_\alpha = \Lambda_R/ \alpha=1/\alpha\,R$ with $\alpha >1$ correspond to
$k_{\rm IR}/k_{\rm eff}<1$ and are acceptable approximations, with the level of precision 
set by $\alpha$ ({\em e.g.}~$k_{\rm IR}/k_{\rm eff} \simeq 0.1$ for $\alpha = 5$).
In particular, we find 
\be
k_{\rm eff}
=
\frac{M^2}
{\mpl^2\,R}
\,f(\alpha)
\ ,
\ee
with $f(\alpha)$ given explicitly in Eq.~\eqref{fa}.
\begin{figure}[t]
\centering
\includegraphics[width=8cm]{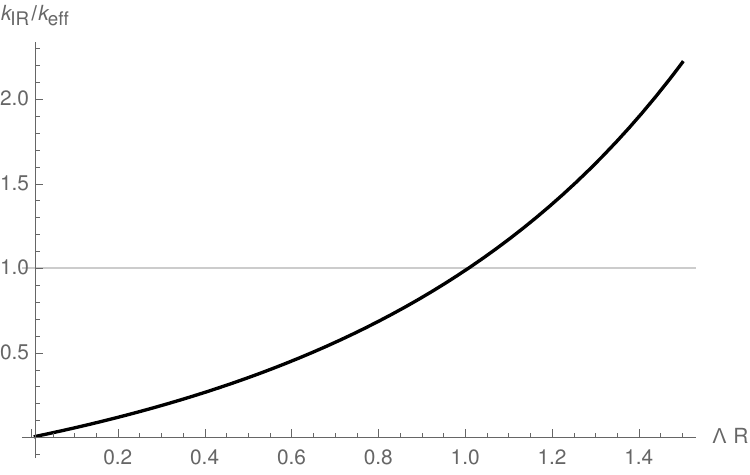}
\caption{Ratio between $k_{\rm IR}$ and $k_{\rm eff}$ for varying $\Lambda$. 
The threshold is $\Lambda_R \simeq 1/R$.}
\label{ratiok}
\end{figure}
\par
\begin{figure}[t]
\centering
\includegraphics[width=8cm]{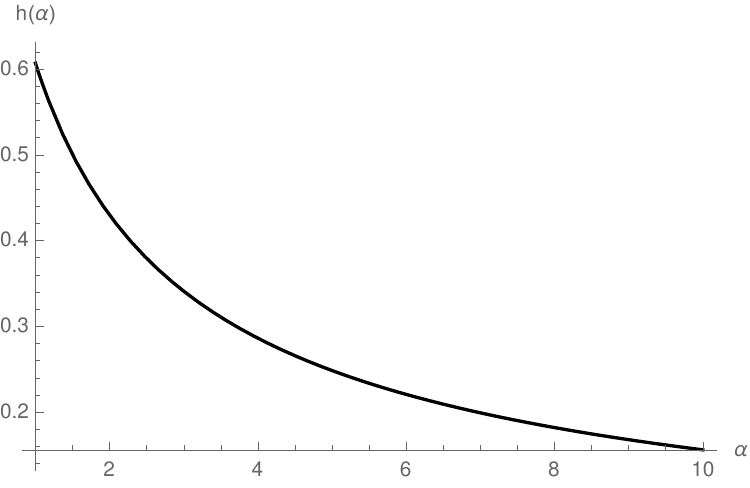}
\caption{Plot of the function $h=h(\alpha)$.
}
\label{halpha}
\end{figure}

We can now use the scale $\Lambda_\alpha$ in order to identify the number $\Nge$ 
of effective (hard) gravitons and the number $N_{\rm G}^{\rm IR}$ of IR gravitons, namely
\be
\Ng
&\!\!=\!\!&
\int_0^{\Lambda_\alpha} 
\frac{d\,k}{2\,\pi ^2} 
\,k^2\,g_k ^2
+
\int_{\Lambda_\alpha}^{\infty} 
\frac{d\,k}{2\,\pi ^2} 
\,k^2\,g_k ^2
 \notag
\\
&\!\!=\!\!&
N_{\rm G}^{\rm IR} + \Nge
\ .
\label{Ngsplit}
\ee
The finite number of gravitons contributing to $k_{\rm eff}\simeq \expec{k}$ is given by
\be
\label{NGeff}
\Nge
=
\frac{M^2}{\mpl^2}\,
g(\alpha)
\ ,
\ee
where $g(\alpha)$ is a numerical factor displayed in Eq.~\eqref{NgNb}.
The infinity (for $R_\infty\to \infty$) in the total amount~\eqref{NgN} comes from $N_{\rm G}^{\rm IR}$,
which counts the very soft gravitons contributing the small $k_{\rm IR}$. 
It is now quite straightforward to evaluate the mean graviton wavelength as
\be
\notag
\lambda_{\rm G}
&\!\!\simeq\!\!&
\frac{\Nge}{k_{\rm eff}}
=
R\,
\frac{f(\alpha)}{g(\alpha)}
\\ \label{lambdaN}
&\!\!\equiv\!\!&
R\,
h(\alpha)
\ .
\ee
Since $h(\alpha)<1$ for $\alpha>1$ (see Fig.~\ref{halpha}), we have
\be
\lambda_{\rm G}(\alpha)
\simeq
h(\alpha)\,
R
\leq
\alpha\,R
\ ,
\ee
and the average wavelength consistently belongs to the effective part of the spectrum
(that is, $1/\lambda_{\rm G}(\alpha)>\Lambda_\alpha$).
\par
We conclude this section by remarking once more that the important results are that 
$\Nge$ only depends on the ADM energy $M$ precisely like in Eq.~\eqref{Ng}, whereas $\lambda_{\rm G}$
is only proportional to $R$, and none of this quantities associated with the coherent state for
the Newtonian potential therefore depend on the compactness of the source.
The corpuscular scaling~\eqref{lambdaRh} for black holes, namely $\lambda_{\rm G}\simeq \Rh\sim M$,
could therefore be obtained only by assuming $R\sim \Rh$.
This all should not be surprising since the Newtonian theory is linear, hence nothing special happens
in it when $R\sim \Rh$ and a black hole is formed.
\section{Bootstrapped gravitational potential}
\label{boot}
\setcounter{equation}{0}
In this section, we briefly recall the definition of the bootstrapped Newtonian gravity described in details
in Refs.~\cite{Casadio:2017cdv,BootN,Casadio:2019cux,Casadio:2019pli}.
In particular, the nonlinear equation for the potential generated by a compact source
is obtained by adding to the Newtonian Lagrangian~\eqref{LagrNewt} several interacting
terms for the field potential $V$.
First of all, we couple $V$ to a gravitational current proportional to its own energy density, 
\be
J_V
=
4\,\frac{\delta U_{\rm N}}{\delta \mathcal{V}} 
=
-\frac{\left[V'(r)\right]^2}{2\,\pi\,\gn}
\ ,
\ee
where $\mathcal{V}$ is the spatial volume and $U_{\rm N}$ the Newtonian potential energy.
The current $J_V$ can also be obtained from the weak field expansion of the Einstein-Hilbert action
to next-to-leading order~\cite{Casadio:2017cdv}.
Moreover, at the same order in that expansion, one finds the term
\be
J_{\rho}
=
-2\,V^2
\ ,
\ee
which couples to the energy density $\rho$.
Finally, since the pressure gravitates and becomes very relevant for large compactness, we add to the
energy density the term~\cite{Casadio:2019cux}
\be
\jb 
=
-\frac{\delta U_{\rm B}}{\delta \mathcal{V}}
\simeq 
p
\ ,
\ee
where $U_{\rm B}$ is the potential energy associated with the work done by the force responsible
for the pressure.
The total Lagrangian then reads
\be
L[V]
&\!\!=\!\!&
L_{\rm N}[V]
-4\,\pi
\int_0^\infty
r^2\,\d r
\left[
q_V\,J_V\,V
+
q_{\rm B}\,J_{\rm B}\,V
+
q_\rho\,J_\rho\left(\rho+p\right)
\right]
\nonumber
\\
&\!\!=\!\!&
-4\,\pi
\int_0^\infty
r^2\,\d r
\left[
\frac{\left(V'\right)^2}{8\,\pi\,\gn}
\left(1-4\,q_V\,V\right)
+V\left(\rho+q_{\rm B}\,p\right)
-2\,q_{\rho} V^2\,\left(\rho+p\right)
\right]
\ ,
\label{LagrV}
\ee
where the coupling constants $q_V$, $q_{\rm B}$ and $q_\rho$ can be used to track the effects of the different
contributions.
Their values would depend on the underlying microscopic quantum theory,~\footnote{See
Refs.~\cite{Casadio:2019cux,Casadio:2019pli} for more details on the role of these coupling parameters.}
but we will here consider only the case $q_V=q_{\rm B}=q_{\rho}=1$ for simplicity, so that the corresponding
field equation for $V$ reads
\be
\triangle V
=
4\,\pi\,\gn\left(\rho+p\right)
+
\frac{2\,\left(V'\right)^2}
{1-4\,V}
\ .
\label{EOMV}
\ee
Finally, one must include the conservation equation 
\be
p'
=
-V'\left(\rho+p\right)
\ .
\label{eqP}
\ee
Eq.~\eqref{EOMV} is understood as the Poisson equation~\eqref{EOMn} with the addition of pressure
and a self-interacting term, while Eq.~\eqref{eqP} is the Newtonian conservation equation which also accounts
for pressure contributing to the matter energy density.
\subsection{Uniform ball}
The above equations were solved in Ref.~\cite{Casadio:2019cux} for a homogeneous ball of matter in vacuum
described by the density~\eqref{HomDens}.
The solutions must satisfy the same regularity conditions~\eqref{bound0} and~\eqref{boundR} of the Newtonian
potential and must approach the Newtonian behaviour far from the source
\be
\label{boundN}
V_{\rm out}(r)
\simeq
V_{\rm N}
=
-\frac{\gn\,M}{r}
\qquad
{\rm for}
\
r\gg R^*
\ ,
\ee
where $M$ is the total ADM energy which is equal to the rest mass $M_0$ only in the Newtonian case.
In general, the relation $M_0=M_0(M)$ is rather involved and is fixed by the boundary conditions~\eqref{boundR}.
The scale $R^*$ introduced above represents a distance (well) beyond which the potential can be safely approximated
by the Newtonian expression in the outer vacuum.
It is therefore natural to identify $R^*$ as the larger between the gravitational radius of the matter source
with energy $M$ and the actual size $R$ of the matter source,
\be
R^* = \max\{\gn\, M, R\}
\ .
\label{Rstar}
\ee
In the following, we will review the (approximate) solutions obtained in Ref.~\cite{BootN,Casadio:2019cux}.
\subsubsection{Outer potential}
In vacuum, where $\rho=p=0$, Eq.~\eqref{EOMV} simplifies to
\be
\triangle V
=
\frac{2 \left(V'\right)^2}{1-4\,V}
\ ,
\label{EOMV0}
\ee
and an exact solution was found in Ref.~\cite{BootN} satisfying the asymptotic condition~\eqref{boundN},
namely
\be
V_{\rm out}
=
\frac{1}{4}
\left[
1-\left(1+\frac{6\,\gn\,M}{r}\right)^{2/3}
\right]
\ .
\label{sol0}
\ee
The right hand sides of Eq.~\eqref{boundR} can therefore be computed exactly,
\be
\label{VR}
V_R
&\!\!=\!\!&
V_{\rm out}(R)
=
\frac{1}{4}
\left[
1-\left(1+\frac{6\,\gn\,M}{R}\right)^{2/3}
\right]
\\ \label{V1R}
V'_R
&\!\!=\!\!&
V_{\rm out}'(R)
=
\frac{\gn\,M}
{R^{2}\left(1+{6\,\gn\,M}/R\right)^{1/3}}
\ ,
\ee
which will be useful in the following.
\subsubsection{Pressure and inner potential}
In the interior of the homogeneous ball, Eq.~\eqref{eqP} can be used to express the pressure as~\cite{Casadio:2019cux}
\be
\label{pressure}
p
=
\rho_0
\left[
e^{V_R-V}
-1
\right]
\ .
\ee
The field equation~\eqref{EOMV} then becomes
\be
\triangle V
=
\frac{3\,\gn\,M_0}{R^3}\,e^{V_R-V}
+
\frac{2\left(V'\right)^2}
{1-4\,V}
\ .
\label{eomV}
\ee
As shown in Ref.~\cite{Casadio:2019cux}, it is possible to find approximate solutions
for $\gn\,M/R \ll 1$ and $\gn\,M/R \gg 1$, which of course give two different relations
between $M$ and $M_0$.
More explicitly, in the low compactness regime $\gn\,M/R \ll 1$, one finds~\footnote{More accurate
approximations can be found in Ref.~\cite{Casadio:2019cux}.}
\be
\label{Vins}
V_{\rm in}
\simeq
V_{\rm s}
=
\frac{\gn\,M}{2\,R}
\left(
1-\frac{2\,\gn\,M}{R}
\right)
\frac{r^2-3\,R^2}{R^2}
\ ,
\ee
with
\be
M_0
&\!\!\simeq\!\!&
\frac{M\,e^{-\frac{\gn\,M}{2\,R\left(1+6\,\gn\,M/R\right)^{1/3}}}}
{\left(1+6\,\gn\,M/R\right)^{1/3}}
\nonumber
\\
&\!\!\simeq\!\!&
M
\left(
1-\frac{5\,\gn\,M}{2\,R}
\right)
\ .
\label{M0s}
\ee
On the other hand, when the compactness is very large, $\gn\,M/R \gg 1$,
the inner solution is well approximated by the linear potential
\be
V_{\rm in}
\simeq
V_{\rm lin}
=
V_R
+
V'_R\left(r-R\right)
\ ,
\label{Vlin}
\ee
and we obtain the relation 
\be
\frac{\gn\,M_0}{R}
\sim
\left(\frac{\gn\,M}{R}\right)^{2/3}
\ ,
\label{M0lin}
\ee
which expresses the compactness in the (hidden) mass $M_0$ in terms of the (observable) compactness 
in the outer mass $M$.
Eq.~\eqref{M0s} also shows that $M_0 \lesssim M$ for $\gn\,M/R \ll 1$,
whereas Eq.~\eqref{M0lin} tells us that $M_0 \ll M$ for $\gn\,M/R \gg 1$.
In both cases the ADM mass is larger than the proper mass of the source.
\section{Scaling relations from the bootstrapped potential}
\label{scaling}
\setcounter{equation}{0}
Everything is set for a quantum interpretation of the bootstrapped potential in terms of a
coherent state following the approach of Section~\ref{coherent}.
Unfortunately, the calculations of the number of gravitons and their mean wavelength are now made
more difficult by the fact that we cannot compute the Fourier transform of the scalar potential $V=V(r)$
and the integrals in $k$ in Eqs.~\eqref{NG0} and~\eqref{kmed0} cannot be done exactly. 
For this reason, we shall employ a different procedure, detailed in Appendix~\ref{calculations},
which amounts to rewriting Eq.~\eqref{kmed0} as the spatial integral~\eqref{eqk}~\footnote{It is crucial that 
the $\Ng$ is still IR divergent while $\expec{k}$ is finite, as shown explicitly in Appendix~\ref{calculations}.},
that is
\be 
\expec{k} 
&\!\!=\!\!&
\frac{2\,\pi}{\lp^2}\int_0^{\infty} \d r\,
r^2\,\left[V'(r)\right]^2
\notag
\\ 
&\!\!=\!\!& 
\frac{2\,\pi}{\lp^2}\int_0^{R} \d r\,
r^2\,\left[V'_{\rm in}(r)\right]^2
+
\frac{2\,\pi}{\lp^2}\int_{R}^{\infty} \d r\,
r^2\,\left[V'_{\rm out}(r)\right]^2
\ ,
\label{khom}
\ee
and then use a similar argument to that of Section~\ref{Newt}.
The main difference is that, since we integrate along the radial coordinate, we must  
determine a length scale $R_{\gamma}$ such that the integral from 0 to $R_{\gamma}$ provides
the main contribution to $\expec{k}$ in Eq.~\eqref{khom}.
\par
We separate the two possible cases with $R_{\gamma}<R$ and $R_{\gamma}>R$, respectively,
and define
\be 
k_{\rm eff}
=
\left\{
\begin{array}{lrl}
\strut\displaystyle
\frac{2\,\pi}{\lp^2}
\int_0^{R_{\gamma}} \d r\,
r^2\,\left[V'_{\rm in}(r)\right]^2
&
{\rm for}
&
0\le  R_{\gamma}<R 
\\
\\
\strut\displaystyle
\frac{2\,\pi}{\lp^2}
\int_0^{R} \d r\,
r^2\,\left[V'_{\rm in}(r)\right]^2
+
\frac{2\,\pi}{\lp^2}
\int_{R}^{R_{\gamma}} \d r\,
r^2\,\left[V'_{\rm out}(r)\right]^2
&
{\rm for}
&
R_{\gamma}>R 
\end{array}
\right.
\label{keff}
\ee
and 
\be
k_{\infty}
=
\left\{
\begin{array}{lrl}
\strut\displaystyle
\frac{2\,\pi}{\lp^2}
\int_{R_{\gamma}}^{R} \d r\,
r^2\,\left[V'_{\rm in}(r)\right]^2
+
\frac{2\,\pi}{\lp^2}
\int_{R}^{\infty} \d r\,
r^2\,\left[V'_{\rm out}(r)\right]^2
&
{\rm for}
&
0\le  R_{\gamma}<R 
\\
\\
\strut\displaystyle
\frac{2\,\pi}{\lp^2}
\int_{R_{\gamma}}^{\infty} \d r\,
r^2\,\left[V'_{\rm out}(r)\right]^2
&
{\rm for}
&
R_{\gamma}>R 
\ .
\end{array}
\right.
\label{kinf}
\ee
The ratio
\be  
\frac{k_{\infty}}{k_{\rm eff}}
=
\gamma
\ ,
\label{ratio}
\ee
with $\gamma <1$, defines the scale $R_{\gamma}$ for which $k_{\rm eff}$ approximates $\expec{k}$
within the required precision (similarly to the parameter $\alpha$ used in Section~\ref{Newt}). 
The analysis in Appendix~\ref{A:NG} shows that the number of gravitons scales as $M^2/\mpl^2$,
under quite general assumptions, and contains the same logarithmic divergence as in the Newtonian case,
with $R^*$ replacing $R$, that is
\be 
\Ng
\simeq
4\, \frac{M^2}{\mpl^2}
\log{\left(\frac{R_\infty}{R^*}\right)}
\ .
\ee 
We shall therefore rely on the argument of Section~\ref{Newt} and assume that the 
number of gravitons effectively contributing up to the scale $R_{\gamma}$ is finite and
proportional to $M^2/\mpl^2$,
\be 
\Nge 
\sim
\frac{M^2}{\mpl^2}
\ .
\label{NeffB}
\ee
In the following, we will estimate the scale $R_{\gamma}$ for the Newtonian potential as a
test of the method and then apply it to the bootstrapped potential.
\subsection{Newtonian potential}
\label{Newtbis}
\begin{figure}[t]
\centering
\includegraphics[width=8cm]{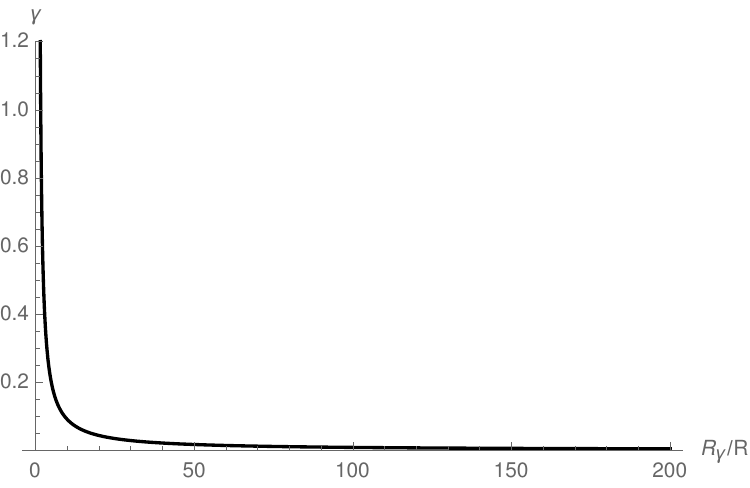}
$\ $
\includegraphics[width=8cm]{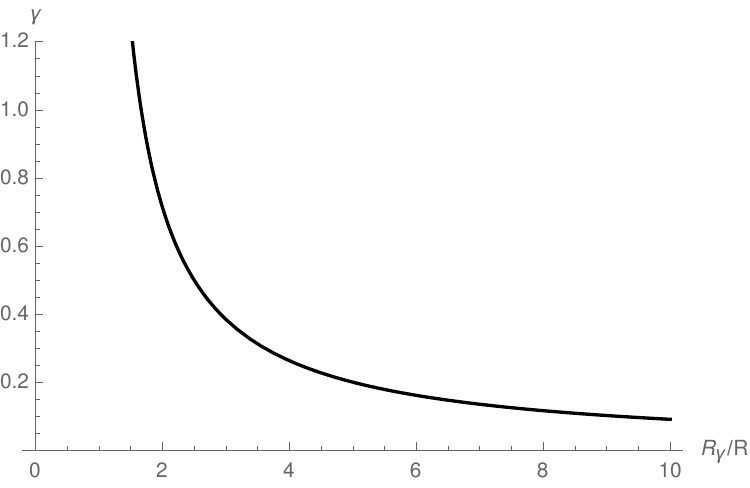}
\caption{Ratio $k_{\infty}/k_{\rm eff}=\gamma$ for the Newtonian potential (left panel)
and a close-up view for small $R_\gamma$ (right panel).}
\label{ratiokN}
\end{figure}

We start with the Newtonian potential in order to test the validity of the above Eqs.~\eqref{khom}, \eqref{keff} 
and~\eqref{kinf}.
The first important check is that Eq.~\eqref{khom} indeed reproduces the result~\eqref{kN}, 
\be 
\expec{k}
&\!\!=\!\!&
\frac{2\,\pi}{\lp^2}\int_0^{R} \d r\,
r^4\,\frac{\gn^2\,M^2}{R^6}
+
\frac{2\,\pi}{\lp^2}
\int_{R}^{R_\infty} \d r\,
\frac{\gn^2\,M^2}{r^2}
\notag
\\
&\!\!=\!\!&
\frac{2\,\pi\,M^2}{5\,\mpl^2\,R}
+
\frac{2\,\pi\,M^2}{\mpl^2\,R}
\notag
\\
&\!\!=\!\!&
\frac{12\,\pi\,M^2}{5\,\mpl^2\,R}
\ .
\ee
It is then easy to verify that Eqs.~\eqref{keff} 
and~\eqref{kinf} give
\be 
k_{\rm eff}
=
\left\{
\begin{array}{lrl}
\strut\displaystyle
\frac{2\,\pi\,M^2\,R_{\gamma}^5}{5\,\mpl^2\,R^6}
&
{\rm for}
&
0\le  R_{\gamma}<R 
\\
\\
\strut\displaystyle
\frac{12\,\pi\,M^2}{5\,\mpl^2\,R}
-
\frac{2\,\pi\,M^2}{\mpl^2\,R_{\gamma}}
&
{\rm for}
&
R_{\gamma}>R 
\end{array}
\right.
\ee
and
\be
k_{\infty}
=
\left\{
\begin{array}{lrl}
\strut\displaystyle
\frac{12\,\pi\,M^2}{5\,\mpl^2\,R}
-
\frac{2\,\pi\,M^2\,R_{\gamma}^5}{5\,\mpl^2\,R^6}
&
{\rm for}
&
0\le  R_{\gamma}<R 
\\
\\
-\strut\displaystyle
\frac{2\,\pi\,M^2}{\mpl^2\,R_{\gamma}}
&
{\rm for}
&
R_{\gamma}>R 
\ .
\end{array}
\right.
\ee
After replacing these expression into Eq.~\eqref{ratio}, it turns out that $\gamma<1$ implies
$R_{\gamma}\gtrsim R$, as shown in Fig.~\ref{ratiokN}.
One can in fact solve Eq.~\eqref{ratio} for $R_{\gamma}$ and find
\be 
R_{\gamma}
=
\frac{5}{6}
\left(
\frac{\gamma +1}{\gamma}
\right)
R
\ .
\label{Rb}
\ee
It would be tempting to set a direct connection with the momentum scale $\Lambda_\alpha$ introduced in 
Section~\ref{Newt} and state that $\Lambda_{\alpha=\gamma} = 1/R_{\gamma}$, but we could not find
a strict proof of this relation.
It is nonetheless reassuring that Eq.~\eqref{Rb} further supports the conclusion that in the Newtonian regime
the only relevant scale for $\expec{k}$ is the radius $R$ of the source.
In any case it is sufficient for our purposes to assume that $\Lambda_{\alpha} = 1/R_{\gamma}$ for
precisions $\gamma\sim\alpha$ and show that the mean wavelength computed with the effective gravitons alone
is qualitatively the same as in Eq.~\eqref{lambdaN}.
\subsection{Bootstrapped potential}
We can finally consider the bootstrap solutions of Section~\ref{boot}. 
When the compactness is small, the solutions in Eq.~\eqref{sol0} and~\eqref{Vins}
follow rather closely the Newtonian behaviour and the results of Section~\ref{Newtbis}
become a very good approximation. 
\par
\begin{figure}[t]
\centering
\includegraphics[width=8cm]{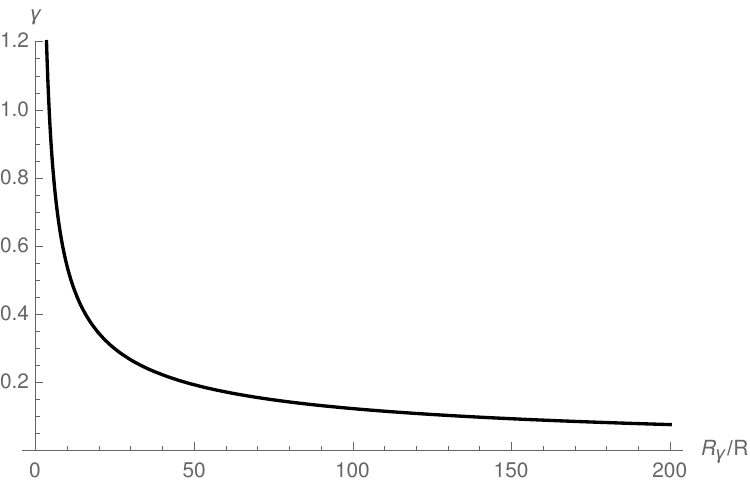}
$\ $
\includegraphics[width=8cm]{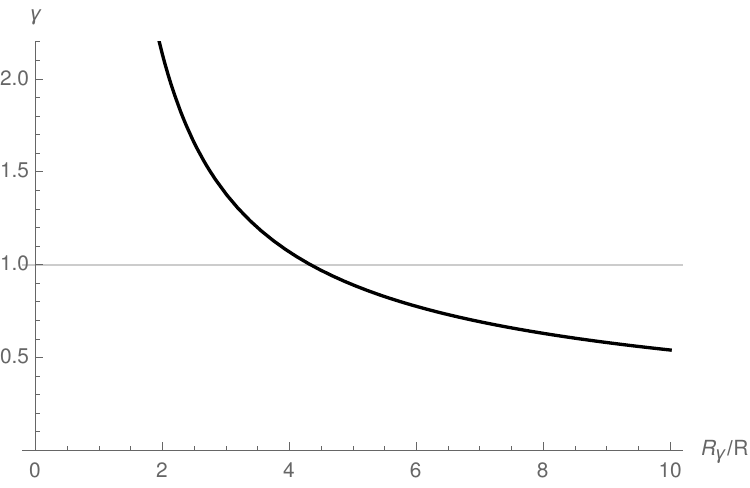}
\caption{Ratio $k_{\infty}/k_{\rm eff}=\gamma$ for the bootstrapped potential (left panel) and close-up
view for small $R_\gamma$ (right panel).}
\label{ratiokB}
\end{figure}
\begin{figure}[t]
\centering
\includegraphics[width=8cm]{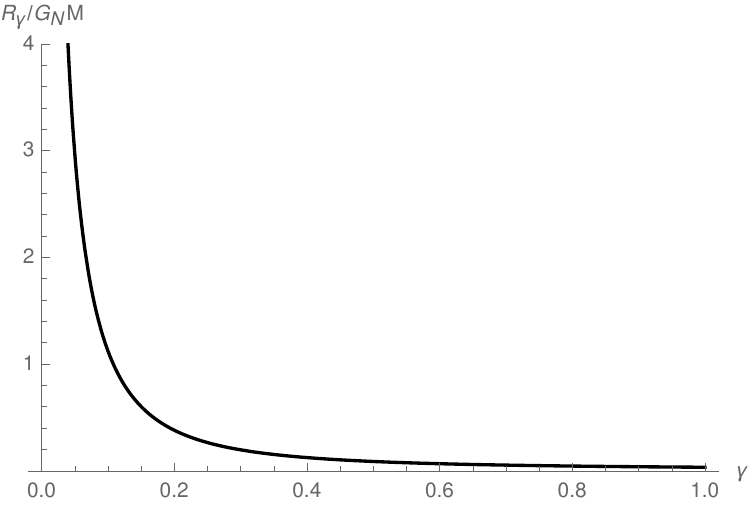}
\caption{$R_{\gamma}$ in units of $\gn\,M$ for the bootstrapped potential.}
\label{Rg}
\end{figure}

When the compactness is instead large, things change significantly.
The outer potential is always given by the exact solution~\eqref{sol0} while
for the inner potential we will consider the linear approximation~\eqref{Vlin}.
In so doing, Eq.~\eqref{khom} gives
\be
\expec{k}
&\!\!\simeq\!\!&
\frac{2\,\pi}{\lp^2}
\int_0^{R} \d r\,
r^2\,\left(V'_{R}\right)^2
+
\frac{2\,\pi}{\lp^2}
\int_{R}^{\infty} \d r\,
r^2\,\left[\frac{\gn\,M}
{\left(1+6\,\gn\,M/r \right)^{1/3}r^2}\right]^2
\notag
\\ 
&\!\!=\!\!&
\frac{2\,\pi\,R^3\,\left(V'_{R}\right)^2}{3\,\lp^2}
+
\frac{2\,\pi\,\gn^2\,M^2}{\lp^2}
\int_{R}^{\infty}
\frac{\d r}
{\left(1+6\,\gn\,M/r \right)^{2/3}r^2}
\notag
\\ 
&\!\!=\!\!&
\frac{\pi\,\gn\,M}{\lp^2}
\left[
\frac{2\,\gn\,M}
{\left(1+{6\,\gn\,M}/R\right)^{2/3}R}
+
\left(1+\frac{6\,\gn\,M}{R}
\right)^{1/3}
-1
\right]
\notag
\\ 
&\!\!\simeq\!\!&
\frac{M}{\lp\,\mpl}
\left(\frac{\gn\,M}{R}
\right)^{1/3}
\ ,
\label{kmedB}
\ee
where $V'_{R}$ is given in Eq.~\eqref{V1R} and the last expression contains just the leading order in the
compactness $\gn\,M/R \gg 1$.
Like in the Newtonian case, the mean wave number $\expec{k}$ is finite, despite the number of gravitons
diverges again and with the same behaviour and functional dependence (see Appendix~\ref{A:NG} for the details).
Given these similarities with the Newtonian regime, we exploit the same method described in 
Section~\ref{Newtbis} in order to find the scale $R_{\gamma}$ for the bootstrapped potentials.
We only consider the case $R_{\gamma}>R$ as it is the only one in which one can have $\gamma<1$.
Hence, Eqs.~\eqref{keff} and~\eqref{kinf} yield
\be
k_{\rm eff}
=
\frac{2\,\pi\,R\left(\gn\,M/R\right)^2}
{3\,\lp^2\left(1+6\,\gn\,M/R\right)^{2/3}}
+
\frac{\pi\,M}{\lp\,\mpl}
\left[
\left(
1+\frac{6\,\gn\,M}{R}
\right)^{1/3}
-\left(
1+\frac{6\,\gn\,M}{R_{\gamma}}
\right)^{1/3}
\right]
\ee
and
\be 
k_{\infty}
=
\frac{\pi\,M}{\lp\,\mpl}
\left[
\left(
1+\frac{6\,\gn\,M}{R_{\gamma}}
\right)^{1/3}
-1
\right]
\ ,
\ee
where the linear approximation~\eqref{Vlin} was considered for the inner potential
and the exact solution~\eqref{sol0} for the outer region.
After solving Eq.~\eqref{ratio} for $R_{\gamma}$, one finds
\be 
R_{\gamma}
\simeq
\frac{6\,\gn\,M}
{\left[
\frac{20}{3\cdot 6^{2/3}}
\left(
\frac{\gamma}{\gamma +1}
\right)
\left(
\frac{\gn\,M}{R}
\right)^{1/3}
+1
\right]^3
-1}
\ .
\label{Rgamma}
\ee
It is easy to see that the threshold value of $R_{\gamma}$, corresponding to $\gamma=1$,
is still proportional to $R$ in the regime $\gn\,M/R\gg 1$.
On the other hand, Figs.~\ref{ratiokB} and~\ref{Rg} show that $R_{\gamma}$ raises very quickly
for $\gamma<1$ and reaches values of order $\gn\,M$ or large for better precisions.
Hence, from Eqs.~\eqref{Rb} and~\eqref{Rgamma}, we see that $R_{\gamma}$ qualitatively behaves
as the scale $R^*$ of Eq.~\eqref{Rstar}: it is proportional to $R$ for sources with small compactness
(consistently with the quasi-Newtonian behaviour) while it is also related to the scale $\gn\,M$
when the compactness becomes large.
In other words, we get a good description of the system by considering gravitons inside a ball
of radius $R_{\gamma}\sim R$ for $\gn\,M/R\ll 1$ and $R_{\gamma}\sim R\,(\gn\,M/R)^{2/3}/\gamma$ for $\gn\,M/R\gg 1$
and $0<\gamma\ll 1$.
In particular, for large compactness, we can tune the precision coefficient $\gamma$ so that
$R_\gamma\sim\gn\,M$.
As we mentioned at the end of Section~\ref{Newtbis}, this suggests that there is a scale $\Lambda \sim 1/R^*$
in momentum space below which the contribution of gravitons becomes essentially irrelevant.
\par
Finally, we simply evaluate the mean graviton wavelength as the ratio between 
Eq.~\eqref{NeffB} and Eq.~\eqref{kmedB} and get
\be
\frac{\lambda_{\rm G}}{R}
\simeq
\left(
\frac{\gn\,M}{R}
\right)^{2/3}
\gg
1
\ ,
\label{lambda}
\ee
so that we can conclude that
\be
1
\lesssim
\frac{\lambda_{\rm G}}{R}
\lesssim
\frac{\gn\,M}{R}
\ ,
\ee
and the compactness of the source yields a (rough) upper bound for the mean wavelength.
The above expression also does not reproduce the expected scaling relation~\eqref{lambdaRh} of the corpuscular
model, to wit $\lambda_{\rm G}\sim M$, unless the compactness is of order one, rather than very large.
However, we will see below that it might be the quantum nature of the source that requires this rather strong
bound for the compactness.
\subsection{Quantum source and GUP for the horizon}
\label{ss:gup}
It was shown in Ref.~\cite{HQMgup} that a quantum source whose size $R$ is comparable with its gravitational
radius~\eqref{Rh} satisfies a GUP~\cite{pGUP} of the form
\be
\Delta R
\sim
\frac{\lp\,\mpl}{\Delta P}
+
\gamma\,\lp\,\frac{\Delta P}{\mpl}
\ ,
\label{hqmGUP}
\ee
where $\Delta R$ is the uncertainty in the size of the source and $\Delta P$ the uncertainty in the conjugate radial momentum.
The first term in the right hand side follows from the usual Heisenberg uncertainty relation, whereas the second term 
corresponds to the horizon fluctuations, $\Delta \Rh\sim\Delta M_0\sim\Delta P$, obtained from the Horizon Wave-Function
(HWF) determining the size $\Rh$ of the gravitational radius~\cite{HQM}.
In Eq.~\eqref{hqmGUP} the two terms are just linearly combined with an arbitrary coefficient $\gamma>0$~\cite{HQMgup}.
In particular, one finds that the quantum fluctuations of the horizon depend strongly on the precise quantum state of the source:
the quantum fluctuations of a macroscopic black hole of mass $M\sim M_0\gg\mpl$ are very large (with $\Delta\Rh/\Rh\sim 1$)
if the source is given by a localised single particle with Compton width $\Delta R\sim R\sim \lp\,\mpl/M_0$~\cite{HQMgup},
whereas they can be negligibly small if the source contains a large number of components of individual energy
$\epsilon\ll M_0$ and size $R\sim\Rh$~\cite{HQMn}, like is the case for corpuscular black holes~\cite{DvaliGomez}.
\par
It is now interesting to note that the relation~\eqref{M0lin} for very compact sources directly implies a similar GUP for the
gravitational radius, namely
\be
\frac{\Delta \Rh}{\Rh}
\simeq
\frac{\Delta M}{M}
&\!\!=\!\!&
\frac{\Delta M_0}{M_0}
+
\frac{\Delta R}{R}
\nonumber
\\
&\!\!\sim\!\!&
\frac{\lp^2}{R^2}
\left(\frac{R}{\gn\,M}\right)^{2/3}
\frac{R}{\Delta R}
+
\frac{\Delta R}{R}
\ ,
\label{dMM}
\ee  
where we again assumed the Heisenberg uncertainty relation for the source,
\be
\Delta M_0 
\sim
\frac{\lp\,\mpl}{\Delta R}
\ ,
\label{dMR}
\ee
and used Eq.~\eqref{M0lin} to express the compactness in terms of the ADM mass $M$.
In particular, the second term in Eq.~\eqref{dMM} is analogous to the second term in Eq.~\eqref{hqmGUP}
and would not be found in the case of Newtonian gravity (where $M=M_0$ exactly), or it would be negligibly small
for small compact sources (for which $M\simeq M_0$).
The fluctuations of the horizon are now dominated by the fluctuations of the source, $\Delta M\sim\Delta R$,
for very large compactness $\gn\,M/R\gg 1$, if the size of the source $R\gtrsim\lp$ (otherwise the usual Heisenberg 
term cannot be neglected). 
This is analogous to the above mentioned results obtained from the HWF
(except for the auxiliary condition $R\gtrsim \lp$).
\par
Let us continue to consider the case of large compactness and note that one needs $\Delta M/M\ll 1$
for the gravitational radius to show a classical behaviour.
This can be obtained for a quasi-classical source with $\Delta R/R\ll 1$ provided the compactness is sufficiently large.
Indeed, we can minimise the above expression~\eqref{dMM}, thus obtaining
\be
\frac{\Delta R}{R}
\simeq
\frac{\lp}{R}
\left(\frac{R}{\gn\,M}\right)^{1/3}
\ .
\ee
The corresponding minimum value of the horizon fluctuations is then given by
\be
\frac{\Delta M}{M}
\simeq
2\,
\frac{\lp}{R}
\left(
\frac{R}{\gn\,M}
\right)^{1/3}
\sim
\frac{\Delta R}{R}
\ ,
\label{minDMM}
\ee
so that the condition of classicality of the source, $\Delta R/R\ll 1$, or
\be
\frac{\gn\,M}{R}
\gg
\frac{\lp^3}{R^3}
\ ,
\ee
seems to ensure that the gravitational radius is also classical and satisfies $\Delta\Rh/\Rh\sim \Delta M/M\ll 1$.
\par
However, the above argument does not yet take into consideration the quantum description of the gravitational potential
in terms of a coherent state.
Indeed, we should note that Eq.~\eqref{lambda} implies that the above minimum uncertainty~\eqref{minDMM} for the horizon
would correspond to a mean graviton wavelength 
\be
\frac{\lambda_{\rm G}}{R}
\sim
\left(
\frac{\gn\,M}{R}
\right)^{2/3}
\sim
\frac{\lp^2}{\Delta R^2}
\ .
\ee
Assuming the matter uncertainty cannot realistically be smaller than the Planck length, this appears to constrain the compactness
to be of order one or less, in clear contradiction with the starting assumption $\gn\,M/R\gg 1$.
On the other hand, for a compactness of order one, both Eq.~\eqref{M0lin} and the analysis of the Newtonian
case in Section~\ref{Newt} would imply that 
\be
\lambda_{\rm G}\sim R\simeq \lp\,\frac{M}{\mpl}
\ ,
\ee
which is precisely the prediction of the corpuscular model~\cite{DvaliGomez}.
Furthermore, we remark that the second approximation in the small compactness expression~\eqref{M0s} clearly
fails for $\gn\,M/R\simeq 1$ and Eq.~\eqref{M0lin} cannot yet be trusted in this intermediate regime~\footnote{We
showed numerically in Ref.~\cite{Casadio:2019cux} that this is in fact the most difficult regime to describe analytically.}.
If we evaluate the first line of Eq.~\eqref{M0s} for $\gn\,M/R\simeq 1$, we obtain
\be
M
\simeq
\frac{3}{2}\,M_0
\ee
and 
\be
\frac{\Delta M}{M}
\simeq
\frac{\Delta M_0}{M_0}
\sim
\frac{\mpl}{M}
\frac{\lp}{\Delta R}
\lesssim
\frac{1}{\sqrt{N_{\rm G}}}
\ ,
\ee
where we used the scaling relation~\eqref{Ng} and $\Delta R/\lp\gtrsim 1$.
This result is consistent with the horizon of a macroscopic black hole (with $N_{\rm G}\gg 1$) being classical.
Finally, we note that the scaling for the fluctuations derived for thermal black holes in Refs.~\cite{HQMn},
\be
\frac{\Delta M}{M}
\sim
\frac{1}{N_{\rm G}}
\ ,
\ee
is recovered from $\Delta R\sim \lambda_{\rm G}\sim \Rh$.
Such a large uncertainty would apply to matter in a truly quantum state, like a condensate or the core
of a neutron star.
\section{Conclusions and outlook}
\label{s:conc}
\setcounter{equation}{0}
In this work, we have investigated the coherent state description for the bootstrapped Newtonian potential
found in Ref.~\cite{Casadio:2019cux} for a uniform spherically symmetric source, and shown that
the scaling relation~\eqref{lambdaRh} for the mean graviton wavelength can be recovered provided the compactness
of the source never exceeds values of order one.
(The similar scaling~\eqref{Ng} for the ADM mass holds regardless.)
Moreover, such a bound on the compactness seems in turn to be required by a consistent quantum description
of both gravity and the matter source itself, so that even macroscopic black holes should be viewed as proper quantum
systems~\cite{DvaliGomez,afshordi}.
We should remark that this result comes with a number of caveats.
\par
First of all, from a quantum field theory perspective, the potential we employ to describe the gravitational 
pull on test particles should emerge from a suitable limit of the interacting propagator for test particles with
the constituents of the matter source.
Considering that we are interested in understanding gravity also in the interior of the self-gravitating object,
and given the complexity of a macroscopic matter source, this approach seems hardly attainable (analytically).
We have therefore assumed that a heuristic description in terms of a scalar potential represents
a sensible mean field approximation, like the Coulomb potential yields a viable quantum description of the 
hydrogen atom or other bound states in quantum electrodynamics.
\par
Another important remark is that, if one views the equation governing the bootstrapped potential as the truncated
version of general relativity, including just the first nonlinearities sounds completely arbitrary and one might
argue that there are no reasons to believe the results would remain unchanged by adding more terms. 
Actually, one could easily argue that, at the classical level, the inclusion of all terms stemming from general
relativity would reintroduce the Buchdahl limit~\cite{buchdahl} and the well-known singularities.
However, if the singularities have to be removed, a modification of general relativity becomes necessary
and the bootstrapped Newtonian potential is just one of the simplest toy models we can employ to study quantum
features of the nonlinear dynamics for macroscopic sources.
On the other hand, if it is indeed possible to recover the (quantum) gravitational dynamics at all orders in perturbation
theory from the leading nonlinearities and diffeomorphism invariance (which is usually referred to as the bootstrap
programme~\cite{rubio}, but see also the approach in Ref.~\cite{hansen}), the results in the present work might help
to understand the gravitational physics of macroscopic matter sources which cannot be treated as small perturbations
about the vacuum.
\par
We would like to conclude with a few more comments and outlook.
It is interesting to notice that the bootstrapped gravitational potential inside very compact sources
being essentially linear shows a similarity with the case of quantum chromodynamics.
Moreover, according to the final result of this work, it appears that the linear regime (analogous to the 
effective gluon potential between two quarks) should never be realised inside static black holes, like quarks
cannot be pulled too far apart but form mesons and hadrons.
We already mentioned in the Introduction that it is tempting to view this picture, in which the compactness
of a self-gravitating object never exceeds values of order one, as pointing to the classicalization~\cite{classicalization,giusti}
in matter-gravity systems.
From the phenomenological point of view, the question naturally arises whether these objects show a proper horizon,
which could have interesting observational consequences for astrophysical black holes
(see, {\em e.g.}~Ref.~\cite{Buoninfante:2020tfb} for a recent proposal).
In order to address this matter and compare directly with observable quantities for black holes in general relativity,
the bootstrapped potential is however not sufficient, and one should first obtain a complete effective metric,
at least in the vacuum outside the source.
This important but complex task is left for future developments.
We once more remark the crucial role of the matter source in supporting this perspective and the importance
of analysing distributions more realistic than the uniform one considered here (see Ref.~\cite{Casadio:2020kbc}
for polytropic stars).
Likewise, the study of both matter and gravitational perturbations about the static solutions will be
essential for understanding the causal structure and possible phenomenological implications of the 
quantum model~\cite{afshordi}.
Finally, we recall that the corpuscular picture of gravity can be applied to cosmology~\cite{cosmo,giusti},
where the Universe is depicted as a cosmological condensate of gravitons and can give rise to dark energy
and dark matter phenomenology~\cite{DM}, and reproduce the Starobinsky model of inflation~\cite{cosmo,inflation}.
It will therefore be very interesting to embed the description of compact sources in bootstrapped Newtonian
gravity within such a cosmological perspective as local impurities affecting the cosmological condensate of gravitons. 
\subsection*{Acknowledgments}
We would like to thank G.~Dvali, A.~Giusti and O.~Micu for useful comments and suggestions.
R.C.~and M.L.~are partially supported by the INFN grant FLAG.
The work of R.C.~has also been carried out in the framework
of activities of the National Group of Mathematical Physics (GNFM, INdAM)
and COST action {\em Cantata\/}. 
%
%
%
%
\appendix
\section{Effective wavenumber and graviton number for the Newtonian potential}
\label{calcfg}
\setcounter{equation}{0}
We show here the explicit calculation of $k_{\rm eff}$ and $\Nge$ for $\Lambda_\alpha = 1/\alpha\,R$
and the corresponding functions $f(\alpha)$ and $g(\alpha)$ of Section.~\ref{Newt}. 
\par
Eq.~\eqref{ksplit} with the $g_k$ given by Eq.~\eqref{gkN} yields
\be 
k_{\rm eff}
&\!\!=\!\!&
\int_{\Lambda_\alpha}^{\infty} 
\frac{\d k}{2\,\pi ^2} 
\,k^3\,g_k ^2
\notag
\\ 
&\!\!=\!\!&
\frac{6\,M^2}{5\,\mpl^2\,R}
\left[
2\,\pi
+
\alpha^3
\left(
3\,\alpha^2
+5
\right)
-
\alpha
\left(
3\,\alpha^4
-{\alpha^2}
+
2
\right)
\cos\!\left(\frac{2}{\alpha}\right)
-\alpha^3
\left(
6\,\alpha +1
\right)
\sin\!\left(\frac{2}{\alpha}\right)
\right.
\notag
\\ 
&&
\left.
\qquad\qquad
-
4\,{\rm Si}\!\left(\frac{2}{\alpha}\right)
\right]
\notag
\\
&\!\!\equiv\!\!&
\frac{M^2}
{\mpl^2\,R}\,
f(\alpha)
\ ,
\label{fa}
\ee
where
\be
{\rm Si}(x)
=
\int_0^x
\d t\,\frac{\sin t}{t}
\label{Si}
\ee
is the sine integral.
Since ${\rm Si}(x\to\infty)=\pi/2$, we correctly obtain that $k_{\rm eff}\to 0$ for $\alpha\to 0$
(that is, for $\Lambda_\alpha\to\infty$).
\par
Likewise, Eq.~\eqref{Ngsplit} with the same $g_k$ of Eq.~\eqref{gkN} reads
\be
\Nge
&\!\!=\!\!&
\int_{\Lambda_\alpha}^{\infty} 
\frac{\d k}{2\,\pi ^2} 
\,k^2\,g_k ^2
\notag
\\ 
&\!\!=\!\!&
\frac{\alpha\,M^2}{2\,\mpl^2}
\left[
3\,\alpha^3
\left(
2\,\alpha^2
+
3
\right)
-
\alpha
\left(
6\,\alpha^4
-
3\,\alpha^2
+
2
\right)
\cos\!\left(\frac{2}{\alpha}\right)
-
\alpha^2
\left(
6\,\alpha^2
+1
\right)
\sin\!\left(\frac{2}{\alpha}\right)
\right.
\notag
\\ 
&&
\left.
\qquad\quad\
-
4\,{\rm Si}\!\left(\frac{2}{\alpha}\right)
\right]
\\
&\!\!\equiv\!\!&
\frac{M^2}
{\mpl^2}\,
g(\alpha)
\ ,
\label{NgNb}
\ee
and we again remark that $\Nge\to 0$ for $\Lambda_\alpha\to \infty$.
\section{Graviton number and mean wavelength for compact sources}
\label{calculations}
\setcounter{equation}{0}
As already pointed out in the main text, the exact analytical calculation of the Fourier transform is not
possible for arbitrary potentials $V=V(\bm{x})$ generated by a compact source.
We will therefore describe here an approximation obtained by rewriting the Fourier transform
$\tilde V=\tilde{V}(\bm{k})$ in terms of a spatial integral of the Laplacian of the scalar field.
In fact, if we apply the Laplacian operator on both sides of Eq.~\eqref{fourierV}, we obtain 
\be 
\tilde{V}(\bm{k})
=
-\frac{1}{k^2} 
\int \d\bm{x}\,
\triangle V(\bm{x})\,
v_{\bm{k}}(\bm{x})
\ .
\ee
\par
Upon substituting the above expression together with Eq.~\eqref{vk} into Eq.~\eqref{NG0} we get
\be 
\Ng 
&\!\!=\!\!& 
\frac{1}{2\,(2\,\pi)^3\lp^2}
\int \d\bm{x}
\int \d\bm{y} \, 
\triangle V(\bm{x})\,
\triangle V(\bm{y}) 
\int \d\bm{k}\,
\frac{e^{i\,\bm{k}\cdot(\bm{x}- \bm{y})}}{k^3}
\notag
\\
&\!\!=\!\!& 
\frac{1}{(2\,\pi)^2\lp^2}
\int_{\mathcal{B}_0^{\infty}}
\d\bm{x}
\int_{\mathcal{B}_0^{\infty}}
\d\bm{y} \, 
\triangle V(\bm{x})\,
\triangle V(\bm{y})
\int_{k_0}^\infty \d k
\, \frac{\sin{(k\,\sigma)}}{k^2\, \sigma}
\ ,
\label{NGr}
\ee
where $\sigma = |\bm{x}-\bm{y}|$ and $k_0=1/R_\infty$ is the IR cut-off introduced in Section~\ref{Newt}
for the purpose of regularising the diverging number of gravitons associated with the infinite spatial
support of the potential.
We have correspondingly restricted the spatial domain of integration to a ball of radius $R_\infty$
centred in the origin, $\mathcal{B}_0^{\infty}=\{|\bm{x}|<R_\infty\}$.
\par
Similarly for the mean wavenumber in Eq.~\eqref{kmed0} we have
\be 
\expec{k} 
&\!\!=\!\!& 
\frac{1}{2\,(2\,\pi)^3\lp^2}
\int \d\bm{x} 
\int \d\bm{y}
\, \triangle V(\bm{x})\, 
\triangle V(\bm{y})
\int \d\bm{k}\, 
\frac{e^{i\,\bm{k}\cdot (\bm{x}- \bm{y})}}{k^2}
\nonumber
\\ 
&\!\!=\!\!& 
\frac{1}{(2\,\pi)^2\lp^2}
\int \d\bm{x} 
\int \d\bm{y} \, 
\triangle V(\bm{x})\, 
\triangle V(\bm{y}) 
\int_0^\infty \d k \, 
\frac{\sin{(k\,\sigma)}}{k\,\sigma}
\nonumber
\\
&\!\!=\!\!& 
\frac{1}{8\,\pi\,\lp^2}
\int \d\bm{x}
\int \d\bm{y} \, 
\frac{\triangle V(\bm{x})\, 
\triangle V(\bm{y})}{\sigma} 
\ ,
\label{kmedr}
\ee
where we used the property of the sine integral~\eqref{Si} that ${\rm Si}(x\to \infty)=\pi/2$. 
This mean wavenumber is regular since only a finite part of the (infinite number of) gravitons
effectively contribute to it, and does not require any cut-off.
\par
Eqs.~\eqref{NGr} and~\eqref{kmedr} show that the divergence of $\Ng$ and 
the finiteness of $\expec{k}$ do not depend on the actual shape of the potential $V$,
as long as it falls off fast enough at large distance.  
We also anticipate that another relevant scale will be given by $R^*$ defined in Eq.~\eqref{Rstar}.
\subsection{Mean graviton wavenumber}
\label{A:kmed}
We will first show how to obtain Eq.~\eqref{khom} from Eq.~\eqref{kmedr}.
This is most easily done if we directly consider a spherically symmetric case such that
\be 
\expec{k} 
=
\frac{1}{8\,\pi\,\lp^2}
\int_0^{\infty} \d r_1 
\int_0^{\infty} \d r_2 \, r_1^2\,r_2^2\, 
\triangle V(r_1)\, \triangle V(r_2) 
\int \d\Omega_1 \int \d\Omega_2 \, 
\frac{1}{|\bm{x} - \bm{y}|}
\ ,
\label{kreg}
\ee
where $\d\Omega_a = \sin{\theta_a}\,\d\theta_a\,\d\varphi_a$, with $a=1,2$.
The freedom to rotate the system allows us to choose $\theta_2$ as the angle between
$\bm{x}$ and $\bm{y}$, which introduces a factor of $8\,\pi^2$ from the integration in
$\d\Omega_1$ and $\d\varphi_2$. 
The only angular integration left is in $\d s\equiv\sin{\theta_2}\,\d\theta_2 = -\d\cos{\theta_2}$,
which yields
\be
\expec{k}
&\!\!=\!\!&
\frac{\pi}{\lp^2}
\int_0^{\infty} \d r_1 
\int_0^{\infty} \d r_2 \, 
r_1^2\,r_2^2\,
\triangle V(r_1)\, 
\triangle V(r_2)
\int_{-1}^1 
\frac{\d s}{\sqrt{r_1^2+r_2^2 +2\,r_1\,r_2\,s}} 
\nonumber
\\
&\!\!=\!\!&
\frac{\pi}{\lp^2}
\int_0^{\infty} 
\d r_1 \int_0^{\infty} 
\d r_2 \, r_1\,r_2\, 
\triangle V(r_1)\, 
\triangle V(r_2) 
\left(r_1+r_2 - |r_1-r_2|\right)
\ .
\ee
Thanks to the  symmetric role of $r_1$ and $r_2$, the above integrals can be written as
\be
\expec{k}
=
\frac{2\,\pi}{\lp^2}
\int_0^{\infty} 
\d r_1\,r_1\,
\triangle V(r_1)
\left[\int_0^{r_1} 
\d r_2\,r_2^2\, 
\triangle V(r_2)
+
r_1\int_{r_1}^{\infty} 
\d r_2\,r_2\, \triangle V(r_2)
\right]
\ .
\ee
From the definition~\eqref{EOMn} of the Laplacian, it is then easy to see that
\be 
\expec{k}
&\!\!=\!\!&
\frac{2\,\pi}{\lp^2}
\int_0^{\infty} \d r_1\,
r_1\,\triangle V(r_1)
\left\{
\int_0^{r_1} \d r_2\,
\frac{\partial}{\partial\,r_2}
\left[r_2^2\,
\frac{\partial\, V(r_2)}
{\partial\,r_2}\right]
+
r_1\int_{r_1}^{\infty} 
\frac{\d r_2}{r_2}
\,\frac{\partial}{\partial\,r_2}
\left[r_2^2\,
\frac{\partial\, V(r_2)}
{\partial\,r_2}\right]
\right\} 
\nonumber
\\ 
&\!\!=\!\!&
\frac{2\,\pi}{\lp^2}
\int_0^{\infty} \d r_1\,
r_1\,\triangle V(r_1)
\left\{
r_1^2\,\frac{\partial\, V(r_1)}{\partial\,r_1}
-
r_1
\left[
r_1\,\frac{\partial\, V(r_1)}{\partial\,r_1}
+
V(r_1)
\right]
\right\}
\nonumber
\\
&\!\!=\!\!&
-\frac{2\,\pi}{\lp^2}
\int_0^{\infty} \d r\,
r^2\,V(r) \,
\triangle V(r)
\ ,
\label{kmed1}
\ee
where we integrated by parts taking into account the boundary conditions~\eqref{bound0} and~\eqref{boundN}.
After integrating by parts again, one finally obtains
\be
\expec{k}
=
\frac{2\,\pi}{\lp^2}
\int_0^{\infty} \d r\,
r^2\,\left[V'(r)\right]^2
\ ,
\label{eqk}
\ee
from which we see that we can indeed estimate $\expec{k}$ directly from the potential $V=V(r)$.
\subsection{Graviton number}
\label{A:NG}
Next, we will show how to estimate $\Ng$ in Eq.~\eqref{NGr}.
Our method relies on the introduction of the characteristic length scale $R^*$ defined in Eq.~\eqref{Rstar} and
in identifying the leading terms in the expansion for large $R_{\infty}/R^*$.
In fact, for the potential generated by a compact source, it is reasonable to consider
$R^* \ll R_{\infty}$, provided the source itself has existed for long enough~\cite{Casadio:2017cdv}. 
\par
\begin{figure}[t]
\centering
\includegraphics[width=8cm]{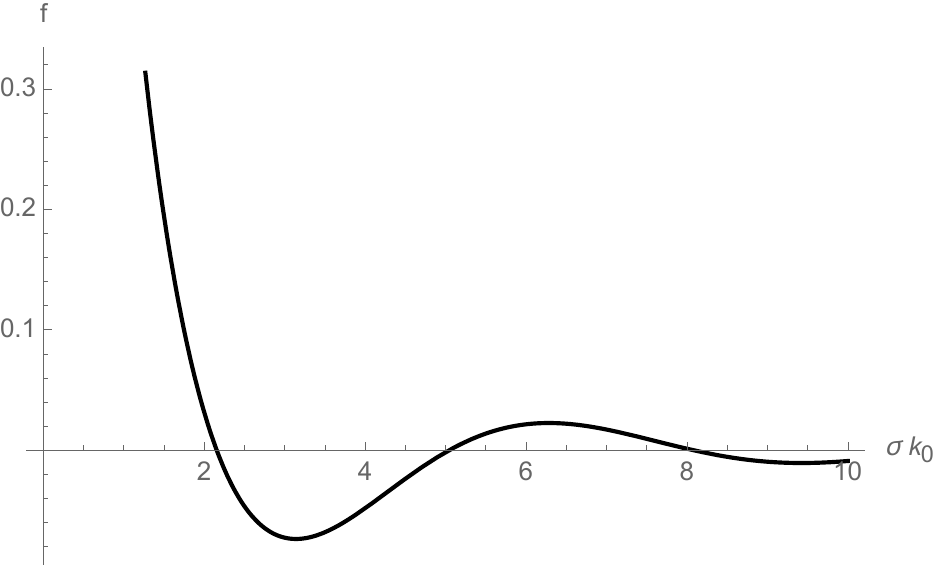}
\caption{Function $f(\sigma)$.}
\label{fsigmaP}
\end{figure}

We first compute explicitly the integral in $k$ in Eq.~\eqref{NGr}, that is
\be
f(\sigma)
&\!\!\equiv\!\!&
\int_{k_0}^\infty
\d k
\, \frac{\sin{(k\,\sigma)}}
{k^2\, \sigma}
\nonumber
\\
&\!\!=\!\!&
\int_{\sigma\,k_0}^\infty
\d z\, 
\frac{\sin{(z)}}{z^2}
\nonumber
\\
&\!\!=\!\!&
\frac{\sin{(\sigma\,k_0)}}{\sigma\,k_0}
-{\rm Ci}(\sigma\,k_0)
\ ,
\label{fsigma}
\ee
where 
\be
{\rm Ci}(x)
&\!\!=\!\!&
\int_0^{x}
\d t\,\frac{1 - \cos{(t)}}{t} 
-\gamma_{\rm E}
-\ln(x)
\ ,
\ee
is the cosine integral and $\gamma_{\rm E}$ the Euler-Mascheroni constant.
It is then easy to show that the function $f(\sigma)$ is larger and contributes significantly
to Eq.~\eqref{NGr} only when its argument $\sigma\ll \Rinf$ (see Fig.~\ref{fsigmaP}).
In fact, for $\sigma \simeq\Rinf$, we have
\be
|f(\sigma)|
\leq 
\int_{\sigma\,k_0}^\infty
\frac{\d z}{z^2} 
= 
\frac{1}{\sigma\,k_0}
=
\frac{\Rinf}{\sigma} 
\simeq 
1
\ .
\ee 
On the other hand, when $\sigma \ll \Rinf$, we can expand Eq.~\eqref{fsigma} for $\sigma\,k_0 \ll 1$,
and note that the leading term is given by $-{\rm Ci}(\sigma\,k_0)\simeq \ln(\sigma\,k_0)$.
To conclude, we can approximate
\be
f(\sigma)
\simeq
\ln{\left(\frac{R_\infty}{\sigma}\right)}
=
\ln{\left(\frac{R_\infty}{R^*}\right)}
+
\ln{\left(\frac{R^*}{\sigma}\right)}
\ ,
\label{fssmall}
\ee
where we explicitly introduced the scale $R^*$.
The second term in Eq.~\eqref{fssmall} diverges for $\sigma = |\bm{x}-\bm{y}| \to 0$, 
but the spatial integrations in Eq.~\eqref{NGr} will regularise it.
In fact, we have explicitly shown in Section~\ref{A:kmed} that the singular function $1/\sigma$
leads to the finite result~\eqref{kreg} once integrated over the spatial domain.
Since $0<-\ln{(\sigma/R^*)}<R^*/\sigma$ for $\sigma\ll R^*$, we can safely neglect the second term in
Eq.~\eqref{fssmall} and just keep the leading contribution coming from the first term
which dominates (and actually diverges) for $\Rinf\gg R^*$.
\par
We must now estimate the spatial integrals in Eq.~\eqref{NGr},
whose domains are effectively restricted by the condition $\sigma=|\bm{x}-\bm{y}| \ll \Rinf$
for which the function $f(\sigma)$ is the largest.
Given the symmetry in $\bm{x}$ and $\bm{y}$, we can achieve this by integrating $\bm{y}$ inside a ball
$\mathcal{B}_{\bm{x}}^{*}$ of radius $R^*\ll \Rinf$ centred around $\bm{x}$ and then summing
over $\bm{x}$ inside ${\mathcal{B}_0^{\infty}}$, that is
\be
\Ng
\simeq
\frac{1}{(2\,\pi)^2\lp^2}
\int_{\mathcal{B}_0^{\infty}}
\d\bm{x} \,\triangle V(\bm{x})
\int_{\mathcal{B}_{\bm{x}}^{*}}
\d\bm{y} \,\triangle V(\bm{y})\,
\log{\left(\frac{R_\infty}{R^*}\right)}
\ .
\ee
The explicit evaluation of this integral is not any simpler than the starting Eq.~\eqref{NGr}.
However, we can now more easily find upper and lower bounds by observing that the Laplacians
are everywhere positive, as can be seen from the fact that the right hand side of Eq.~\eqref{EOMV}
is positive.
An upper bound is obtained by extending the domain of $\bm{y}$ to all of $\mathcal{B}_0^{\infty}$,
\be 
\Ng
&\!\!\leq\!\!&
\frac{1}{(2\,\pi)^2\lp^2}
\int_{\mathcal{B}_0^{\infty}}
\d\bm{x} \,\triangle V(\bm{x})
\int_{\mathcal{B}_0^{\infty}}
\d\bm{y} \,\triangle V(\bm{y})\,
\log{\left(\frac{R_\infty}{R^*}\right)}
\notag
\\
&\!\!\simeq\!\!&
4\, \frac{M^2}{\mpl^2}
\log{\left(\frac{R_\infty}{R^*}\right)}
\ ,
\ee
where we used the Gauss theorem in the form
\be  
\int_{\mathcal{B}_0^{\infty}}
d\bm{x}\,\triangle V(\bm{x})
&\!\!=\!\!&
\int_{\partial \mathcal{B}_0^{\infty}}
d\bm{s}\cdot
\bm{\nabla} V
\nonumber
\\
&\!\!\simeq\!\!&
R_\infty^2 \int d\Omega\, 
\frac{\gn\,M}{R_\infty^2}
\nonumber
\\
&\!\!\simeq\!\!&
4\pi\, \gn\, M
\ ,
\label{gauss}
\ee
with $\d\bm{s} = R^{2}_{\infty}\,\d\Omega\,\bm{n}$ the measure on the sphere $\partial \mathcal{B}_0^{\infty}$
of radius $\Rinf$ whose unit normal vector is $\bm{n}$.
Note also that the second line follows from the Newtonian behaviour at large distance from the source,
namely for $r\gtrsim R^*$.
A lower bound can be obtained by first restricting the domain of $\bm{x}$ to a ball $\mathcal{B}_{0}^{*}$ of radius $R^*$
and then, instead of integrating $\bm{y}$ over all the balls centred around $\bm{x}$, only taking the one centred
in the origin as well. 
The result is
\be
\Ng
&\!\!\geq\!\!&
\frac{1}{(2\,\pi)^2\lp^2}
\int_{\mathcal{B}_0^{*}}
\d\bm{x} \,\triangle V(\bm{x})
\int_{\mathcal{B}_{\bm{x}}^*}
\d\bm{y} \,\triangle V(\bm{y})\,
\log{\left(\frac{R_\infty}{R^*}\right)}
\notag
\\
&\!\!\geq\!\!&
\frac{1}{(2\,\pi)^2\lp^2}
\int_{\mathcal{B}_0^{*}}
\d\bm{x} \,\triangle V(\bm{x})
\int_{\mathcal{B}_0^{*}}
\d\bm{y} \,\triangle V(\bm{y})\,
\log{\left(\frac{R_\infty}{R^*}\right)}
\\
&\!\!\simeq\!\!&
4\, \frac{M^2}{\mpl^2}
\log{\left(\frac{R_\infty}{R^*}\right)}
\ ,
\ee
where we used the defining assumption of $R^*$ that
\be
V'(R^*)
\simeq
\frac{\gn\,M}{(R^*)^2}
\ .
\ee
Therefore, we can safely approximate $\Ng$ as
\be \label{NG}
\Ng 
\simeq
4\, \frac{M^2}{\mpl^2}
\log{\left(\frac{R_\infty}{R^*}\right)} 
\ .
\ee
We point out that this result only depends on the boundary conditions on the potential
at large distance from the source and bares no dependence on the details of the source
or of the gravitational interaction at shorter distances.
\par 
We conclude by estimating the number of effective gravitons.
Like in Section~\ref{Newt}, we introduce the splitting scale $\Lambda$ in Eq.~\eqref{fsigma} and write
\be
f(\sigma)
&\!\!=\!\!&
\int_{\sigma\,k_0}^{\sigma\,\Lambda}
\d z \, 
\frac{\sin{(z)}}{z^2}
+
\int_{\sigma\,\Lambda}^{\infty}
\d z \, 
\frac{\sin{(z)}}{z^2}
\notag
\\
&\!\!=\!\!&
f^{\rm IR}
+
f^{\rm eff}
\ ,
\ee
where $f^{\rm IR}$ is dominated by the logarithmic IR divergence in Eq.~\eqref{fssmall} for $k_0=1/\Rinf\to 0$.
For the finite part, we obtain 
\be 
f^{\rm eff}
=
\frac{\sin{(\sigma\,\Lambda)}}
{\sigma\,\Lambda}
+
\int_0^{\sigma\,\Lambda}
\d t\, \frac{1 - \cos{(t)}}{t} 
- 
\gamma_{\rm E} 
- 
\ln{\left(\sigma\,\Lambda\right)}
\ ,
\ee
in which the dominant term is again given by $\ln{\left(\sigma\,\Lambda\right)}$
for $\sigma\,\Lambda$ small (but still larger then $\sigma\,k_0$).
Since again $0<-\ln{(\sigma\,\Lambda)}<1/\sigma\,\Lambda$, we obtain
\be
\Ng
&\!\!\lesssim\!\!&
\frac{1}{(2\,\pi)^2\lp^2\,\Lambda}
\int \d\bm{x}
\int \d\bm{y} \, 
\frac{\triangle V(\bm{x})\, 
\triangle V(\bm{y})}{\sigma} 
\\
&\!\!\simeq\!\!&
\frac{\expec{k}}{\Lambda}
\ .
\ee
In Section~\ref{scaling}, we show that we can consider $\Lambda \sim 1/R^*$, from
which we obtain for the mean wavelength
\be
\lambda_{\rm G}
\simeq
\frac{\Nge}{\expec{k}}
\lesssim
R^*
\ ,
\ee
so that again this representative scale belongs to the effective part of the spectrum,
that is $1/\lambda_{\rm G}\gtrsim\Lambda$.
\end{document}